\documentclass[12pt]{article}

 \usepackage{amssymb}
 \usepackage{amsmath}
 \usepackage{amsfonts}
 \newcommand{\be}{\begin{equation}}
 \newcommand{\ee}{\end{equation}}
 \newcommand{\bea}{\begin{eqnarray}}
 \newcommand{\eea}{\end{eqnarray}}
 \newcommand{\nn}{\nonumber}
 \newcommand{\rd}{\partial}

\begin{document}

 \begin{titlepage}
  \thispagestyle{empty}

  \vspace{2cm}

  \begin{center}
    \font\titlerm=cmr10 scaled\magstep4
    \font\titlei=cmmi10 scaled\magstep4
    \font\titleis=cmmi7 scaled\magstep4
     \centerline{\titlerm 5D Extremal Rotating Black Holes and CFT duals }

    \vspace{1.5cm}
    \noindent{{%\large
        Farhang Loran\footnote{e-mail: loran@cc.iut.ac.ir},
        Hesam Soltanpanahi\footnote{e-mail: h\_soltanpanahi@ph.iut.ac.ir}
         }}\\
    \vspace{0.8cm}

   {\it Department of  Physics, Isfahan University of Technology,  \\
  Isfahan 84156-83111,  Iran}

  \end{center}

  \vskip 2em

  \begin{abstract}
   Kerr/CFT correspondence has been recently applied to
 various types of 5D extremal rotating black holes.
 A common feature of all such examples is the existence of two chiral CFT duals
 corresponding to the U(1) symmetries of the near horizon geometry.
 In this paper, by studying the moduli space of the near horizon metric of five
 dimensional extremal black holes which are asymptotically flat or
 AdS, we realize an SL(2,${\mathbb Z}$) modular group which is
 a symmetry of the near horizon geometry.
 We show that there is a lattice of chiral CFT duals corresponding to the moduli points identified under the action of the modular group.
 The microscopic entropy corresponding to
 all such CFTs are equivalent and are in agreement with the Bekenstein-Hawking entropy.
  \end{abstract}

\end{titlepage}

 \tableofcontents

 %\keywords{Black Holes in String Theory, Supergravity Models }

%----------------------------------------------------------------------------------------------------------------
 \section{Introduction }\label{int}
 The statistical mechanical interpretation of the Bekenstein-Hawking entropy of black holes
 seems to be an everlasting source of inspiration in quantum gravity.
 The work of Strominger and Vafa \cite{SV} which is known as
 the first attempt to calculate entropy of black holes by counting
 the corresponding microstates  implies that, in principle any gravity solution should have various CFT duals.

 A recent approach in this direction  is the Kerr/CFT correspondence \cite{GHSS},
 which is intrinsically a generalization of Brown-Henneaux approach \cite{BH, strominger} to AdS/CFT correspondence  \cite{C, S, P}.
  Extremal Kerr-AdS metrics in 5, 6 and 7 dimensions are studied in
 \cite{LMP}, and BMPV black hole is studied in \cite{ITW}. CFT dual of 5D extremal rotating Kaluza-Klein black holes
 is investigated in \cite{AOT}.
 Some examples of extremal black hole/CFT correspondence in 4D and 5D  gauged and ungauged
 supergravities are considered in \cite{CCLP}.
 Embedding of such correspondence in string theory for some 5D solutions is studied
 in \cite{N, AOT1}.
 All of these calculations are based on the periodicity  of the circles
 of the near horizon metric and enhancement of the corresponding $U(1)$ symmetries to the  Virasoro of a dual chiral CFT.
 By choosing special boundary conditions, it is shown that there is a CFT dual associated to each of the
 circles, with the same CFT entropy that is in agreement with the Bekenstein-Hawking entropy.

 In this paper we study the moduli space of the near horizon metric of five dimensional extremal rotating black holes
 that are asymptotically flat or AdS. We show that there is an SL(2,${\mathbb Z}$) modular group
 generated by the usual transformations $\tau\to\tau+1$ and $\tau\to -1/\tau$,
 where $\tau$ is a definite moduli of the geometry,
 that leaves the Bekenstein-Hawking entropy invariant.
 The operation  $\tau\to-1/\tau$ interchanges the two circles of the near horizon geometry,
 so maps the  CFT's corresponding to these circles into each other.
 The operator $\tau\to\tau+1$ leaves one of the circles intact but adds it once to the other circle.
 These operations are symmetries of the near horizon geometry but map a given CFT dual to another
 CFT.
 Thus the SL(2,${\mathbb Z}$) modular group assigns a lattice of CFT duals
 corresponding to the moduli points identified under the action of the modular group.
 By using the Cardy formula we  show that the CFT entropy of all of these CFTs are equivalent  and are in
 precise agreement with the Bekenstein-Hawking entropy.

 The organization of this paper is as follows.
 In section \S\ref{gnh}, we study the moduli space of  the  near
 horizon geometry of 5D extremal rotating black holes that are
 asymptotically flat or AdS.
 In section \S\ref{asg} we use the Brown-Henneaux method to define
 the Virasoro generators in terms of the asymptotic symmetry group of
 near horizon geometry and compute the associated central charge.
 In section \S\ref{te} by studying the behavior of the matter field near the horizon of the black hole we compute  the
 Frolov-Thorne (FT) temperature \cite{FT}  and the CFT entropy for the lattice of the CFT
 duals. The consistency of boundary conditions is shown in section
 \S\ref{app}.

 %----------------------------------------------------------------------------------------------------------------
 \section{Extremal 5D Rotating Black Holes  }\label{gnh}
 In the Brown-Henneaux method \cite{BH, strominger}, the first step to construct the CFT dual to a geometry,
  is to realize the metric of the near horizon geometry \cite{GHSS, LMP, ITW, AOT, CCLP, N, AOT1}.
 The general form of near horizon of rotating black
 holes in  5D  that are asymptotically   flat
 or AdS is given in \cite{KLR}. In this paper we follow the conventions of
 \cite{CCLP},
 \be
 ds^2=-(1+\hat{r}^2l^{-2})d\hat{t}^2+\frac{d\hat{r}^2}{1+\hat{r}^2l^{-2}}+
 \hat{r}^2(d\theta^2+\cos^2\theta d\hat{\phi}_1^2+\sin^2\theta
 d\hat{\phi}_2^2),
 \ee
 where $l^{-2}$ is the cosmological constant.
 For deriving the near horizon geometry of the extremal limit, it is suitable to use the following coordinate
 transformations \cite{CCLP}
 \be
 \hat{r}=r_0(1+\varepsilon r),\hspace{7mm}
 \hat{\phi}_1=\phi_1+\Omega_1^0\hat{t},\hspace{7mm}
 \hat{\phi}_2=\phi_2+\Omega_2^0\hat{t},\hspace{7mm}
 \hat{t}=\frac{t}{2\pi r_0T^{'0}_H\varepsilon},\label{hat}
 \ee
 where $r_0$ is the radius of horizon in the extremal limit,
 $\Omega_i^0$ are the angular velocities on the horizon, and the quantity  $T^{'0}_H$
 is derivative of Hawking temperature with respect to the outer horizon radius  at
 itself,
 \be
 T^{'0}_H:=\frac{\rd T_H}{\rd r_+}\Big|_{r_+=r_0}.
 \ee
 The  near horizon metric of the extremal  five dimensional black
 hole is obtained in the  $\varepsilon\rightarrow0$, which gives
 \be
 \begin{array}{l}
 ds^2=A(\theta)\left(-r^2dt^2+\frac{dr^2}{r^2}\right)+F(\theta)d\theta^2+B_1(\theta)(e_1+C(\theta)e_2)^2+B_2(\theta)e_2^2,\\
 e_1=d\phi_1+k_1rdt,\hspace{20mm}e_2=d\phi_2+k_2rdt,
 \end{array}
 \label{met}
 \ee
 where $A, B_i, C$ and $F$ are only functions of $\theta$ ,  $0\leq\phi_i\leq2\pi$ and $0\leq\theta\leq\pi$.
 This metric can be viewed as an S$^3$ bundle over AdS$_2$.
 The constants $k_i=(2\pi T_i)^{-1}$,
 in which $T_i$ are the FT temperatures  defined as follows
 \cite{CCLP},
 \be
 T_i=\lim_{r_+\rightarrow r_0}\frac{T_H}{\Omega_i^0-\Omega_i}
 =-\frac{T^{'0}_H}{\Omega^{'0}_i}.\label{Temp}
 \ee
 Thus the parameter space of asymptotically flat or AdS 5D black holes,
 is seven dimensional with parameters $A, B_i, C, F, k_i$.

 The last two terms of the metric (\ref{met}) can be realized as
 a torus
 \be
 ds^2_{{\mathbb T}^2}=(R_1)^2|e_1+\tau e_2|^2,\label{metT2}
 \ee
 where
 \be
 R_1=\sqrt{B_1},\hspace{15mm}\tau=C+i\sqrt{\frac{B_2}{B_1}}.\label{T2}
 \ee
 The generators of the  SL(2,${\mathbb Z}$) modular group of
 ${\mathbb T}^2$ are
 \be
 \tau\to -\frac{1}{\tau},\hspace{6mm}\tau\to\tau+1.
 \ee
 The action of the modular group of the torus can be extended to the
 parameter space of the near horizon geometry (\ref{met}) as follows
 \bea
 &&(C, B_1, B_2, k_1, k_2)\to(-C|\tau|^{-2}, B_1|\tau|^2, B_2|\tau|^{-2}, -k_2, k_1),\label{mg1}\\
 &&(C, B_1, B_2, k_1, k_2)\to(C+1, B_1, B_2, k_1-k_2, k_2),\label{mg2}
 \eea
 with $A$ and $F$ invariant.
 The corresponding  change in the metric (\ref{met}) can be compensated by a redefinition of the $\phi_i$ coordinates
 given by
 \bea
 \label{revised1}
 &&\phi_1\rightarrow\tilde\phi_1=-\phi_2,\hspace{14mm}\phi_2\rightarrow\tilde\phi_2=\phi_1,\\
 \label{revised2}
 &&\phi_1\rightarrow\tilde\phi_1=\phi_1-\phi_2,\hspace{8.2mm}\phi_2\rightarrow\tilde\phi_2=\phi_2.
 \eea
  Thus, the  SL(2,${\mathbb Z}$) modular transformations (\ref{mg1}) and (\ref{mg2}) are symmetries of the near
  horizon geometry (\ref{met}).\footnote{By transformations (\ref{revised1}) and (\ref{revised2}),
  the orientation and the periods of  the near horizon geometry do not change.}
 Consequently, the Bekenstein-Hawking entropy for the extremal black hole
 \be
 S_{BH}=\frac{1}{4}\int d\theta
 d\phi_1d\phi_2\sqrt{F(\theta)B_1(\theta)B_2(\theta)},\label{BH}
 \ee
 is invariant under the SL(2,${\mathbb Z}$) modular group.
 Here and henceforth we set $G_5=1$.

 %----------------------------------------------------------------------------------------------------------------
 \section{ Asymptotic Symmetry Group}\label{asg}

 The asymptotic
 symmetry group (ASG) of a near horizon metric is the group of allowed
 symmetries modulo trivial symmetries. By definition, an allowed symmetry transformation
  obeys the specified  boundary conditions \cite{GHSS}.
 A possible boundary condition for the fluctuations around the geometry
 (\ref{met})   is,
 \be
 h_{\mu\nu}\sim\mathcal{O}\left(\begin{array}{ccccc}
 r^2&~ 1/r^2&~ 1/{r}&~ r &~ r \\
 ~&~{1}/{r^3}&~ {1}/{r^2}&~{1}/{r}&~{1}/{r}\\
 ~&~&~{1}/{r}&~{1}/{r}&~{1}/{r}\\
 ~&~&~&~{1}&~1\\
 ~&~&~&~&~1
 \end{array}\right),\label{h}
 \ee
 in the basis $(t, r, \theta, \phi_1, \phi_2)$.
 This boundary conditions are consistent with the SL(2,${\mathbb
 Z}$) symmetry of the near horizon geometry given in (\ref{met}).
 This should be contrasted against  the
  boundary conditions,
  \bea
 h_{\mu\nu}^{(1)}\sim\mathcal{O}\left(\begin{array}{ccccc}
 r^2&~ 1/r^2&~ 1/{r}&~ 1 &~ r \\
 ~&~{1}/{r^3}&~ {1}/{r^2}&~{1}/{r}&~{1}/{r^2}\\
 ~&~&~{1}/{r}&~{1}/{r}&~{1}/{r}\\
 ~&~&~&~{1}&~1\\
 ~&~&~&~&~1/r
 \end{array}\right),\label{h1}\\
 h_{\mu\nu}^{(2)}\sim\mathcal{O}\left(\begin{array}{ccccc}
 r^2&~ 1/r^2&~ 1/{r}&~ r &~ 1 \\
 ~&~{1}/{r^3}&~ {1}/{r^2}&~{1}/{r^2}&~{1}/{r}\\
 ~&~&~{1}/{r}&~{1}/{r}&~{1}/{r}\\
 ~&~&~&~{1}/{r}&~1\\
 ~&~&~&~&~1
 \end{array}\right),\label{h2}
 \eea
 considered in  \cite{LMP, ITW, AOT, CCLP, N, AOT1} which are not consistent with the
  modular transformation $\tau\rightarrow\tau+1$ and  by $\tau\to
  -1/\tau$ get mapped into each other.

 It is easy to show that the general diffeomorphism preserving the boundary conditions (\ref{h})
 is given by,
 \bea
 \zeta&=&\left[C+\mathcal{O}(\frac{1}{r^3})\right]\rd_t+\left[r\epsilon(\phi_1,\phi_2)+\mathcal{O}(1)\right]\rd_r+
 \mathcal{O}(\frac{1}{r})\rd_\theta\nn\\
 &+&\left[\lambda_1(\phi_1,\phi_2)+\mathcal{O}(\frac{1}{r^2})\right]\rd_{\phi_1}+
 \left[\lambda_2(\phi_1,\phi_2)+\mathcal{O}(\frac{1}{r^2})\right]\rd_{\phi_2},\label{solution}
 \eea
 where $\epsilon(\phi_1, \phi_2)$ and $\lambda_i(\phi_1, \phi_2)$  are arbitrary  smooth periodic functions
 of $\phi_1$ and $\phi_2$.
 The traceless condition for $h_{\mu\nu}$ indicates that
 \be
 \epsilon(\phi_1,\phi_2)+\rd_{\phi_1}\lambda_1(\phi_1,\phi_2)+\rd_{\phi_2}\lambda_2(\phi_1,\phi_2)=0.\label{constraint}
 \ee
 Thus the ASG contains a class of  generators
 \be
 \zeta=\lambda_1(\phi_1,\phi_2)\rd_{\phi_1}+\lambda_2(\phi_1,\phi_2)\rd_{\phi_2}-
 r\left[\rd_{\phi_1}\lambda_1(\phi_1,\phi_2)+\rd_{\phi_2}\lambda_2(\phi_1,\phi_2)\right]\rd_r.
 \label{revised3}
 \ee
 A subalgebra of the generators (\ref{revised3}) is a Virasoro
 algebra $[\zeta_m , \zeta_n]_{\rm
 Lie}=-i(m-n)\zeta_{m+n}$, where
 \be
 \zeta_m=-e^{-im\phi_1}\rd_{\phi_1}-e^{-im\phi_2}\rd_{\phi_2}-
 imr(e^{-im\phi_1}+e^{-im\phi_2})\rd_r,\label{zm}
 \ee
 corresponding to  $\lambda_i(\phi_1,\phi_2)=\lambda_i(\phi_i)$,
 $i=1,2$ in Eq.(\ref{revised3}).
 %----------------------------------------------------------------------------------------------------------------
 \subsection{Central Charge}\label{cc}
 Charges associated to the diffeomorphisms
 (\ref{solution}) are defined by \cite{BB, BC},
 \be
 Q_\zeta=\frac{1}{8\pi}\int_{\rd\Sigma}k_{\zeta}[h,g],\label{Q}
 \ee
 where $\rd\Sigma$ is spatial surface at infinity and
 \bea
 k_\zeta[h,g]&=&\frac{1}{2}[\zeta_\nu\nabla_\mu h-\zeta_\nu\nabla_\sigma h_\mu^{~\sigma}+\zeta_\sigma\nabla_\nu h_\mu^{~\sigma}+
 \frac{h}{2}\nabla_\nu\zeta_\mu\nn\\&-&h_\nu^{~\sigma}\nabla_\sigma\zeta_\mu
 +\frac{1}{2}h_{\nu\sigma}(\nabla_\mu\zeta^\sigma+\nabla^\sigma\zeta_\mu)]*(dx^\mu\wedge
 dx^\nu),
 \eea
  in which  $*$ denotes the Hodge dual in 5D.
 Since we are interested in the solution which remains extremal
 we  set $Q_{\rd_t}$ to zero.

 In the Brown-Henneaux  approach \cite{BH}
  the central charge is given by
 \be
 \frac{1}{8\pi}\int_{\rd\Sigma}k_{\zeta_m}[\mathcal{L}_{\zeta_n},g]
 =-\frac{i}{12}c(m^3- m)\delta_{{m+n},0}.\label{Qc}
 \ee
 For the metric (\ref{met}) and diffeomorphisms (\ref{zm}),   the central charge is
 \be
 c=\frac{3(k_1+k_2)}{2\pi}\int d\theta
 d\phi_1d\phi_2\sqrt{F(\theta)B_1(\theta)B_2(\theta)}=\frac{6(k_1+k_2)S_{BH}}{\pi}.\label{c-phi}
 \ee
 That is, $c=c_1+c_2$ where $c_i=6\pi^{-1}k_iS_{BH}$, $i=1,2$, are the central charges  of the two CFTs
  associated to the $\phi_i$ circles in \cite{CCLP}.
  This result is in agreement with the central charge
  $c=c_{grav}+c_{gauge}$ assigned to the four-dimensional Kerr-Newmann-AdS-dS
  black hole viewed as a 5D solution \cite{HMNS}.

 In general, the central charge of each point in the lattice of CFT duals defined by Eqs.(\ref{mg1}) and (\ref{mg2})
 is given by
 \be
 c_{(m, n)}=\frac{6(mk_1+nk_2)S_{BH}}{\pi},
 \label{mn}
 \ee
 where $m=a+c$ and $n=b+d$ in which,
 \be
 \left(\begin{array}{cc}a&b\\c&d\end{array}\right)\in SL(2,{\mathbb
 Z}).
 \ee

 Notice that, the SL(2,${\mathbb Z}$) modular symmetry is absent if one of the $k_i$ is zero (like supersymmetric 5D black ring)
  since the  components $g_{t \phi_i}$
 are not of the same order in $r$. For such cases the central charge in the Brown-Henneaux
 approach is expected to be equivalent to the central charge that
 can be calculated in the usual AdS/CFT approaches.
 This is verified for the  supersymmetric 5D black ring in \cite{LSNHBR}.

 %----------------------------------------------------------------------------------------------------------------
 \section{Frolov-Thorne Temperature and Entropy}\label{te}
 The FT temperature can be determined by identifying  quantum numbers
 of a matter field in the near horizon geometry with those in
 original geometry \cite{LMP}.
 For the chiral CFT given by (\ref{zm}) a  matter field can be expanded in
 eigen modes of the asymptotic energy $\omega$
 and {\em angular momentum} $m$ as
 \be
 \Phi=\sum_{\omega , m , l}\varphi_{\omega m l} e^{-i(\omega t-m\phi_+)}f_{l}(r,\theta),\hspace{5mm}\phi_\pm=\phi_1\pm\phi_2,\label{Phi}
 \ee
 since $[\zeta_0,\phi_-]=0$.
 Using (\ref{hat}), the identification
 \be
 e^{-i\omega\hat{t}+i m\hat{\phi}_+}=e^{-in_tt+in\phi_+},
 \ee
  implies that
 \be
 n=m,\hspace{7mm} \omega=2\pi r_0T^{'0}_Hn_t+(\Omega^0_1+\Omega^0_2)n.
 \ee
 Considering the Boltzmann factor,
 \be
 e^{-\frac{\omega-m(\Omega_{1}+\Omega_{2})}{T_H}}=
 e^{-(\frac{n_t}{T_t}+\frac{n}{T})},
 \ee
 one obtains
 \be
 \frac{1}{T_t}=\frac{2\pi r_0T^{'0}_H}{T_H}\hspace{10mm}
 \frac{1}{T}=
 \frac{(\Omega^0_1-\Omega_1)+(\Omega^0_2-\Omega_2)}{T_H}.
 \ee
  It is easy to show that at the extremal limit we have
 \be
 T_t=0,\hspace{15mm} \frac{1}{T}=\frac{1}{T_1}+\frac{1}{T_2},\label{T}
 \ee
 where $T_i=(2\pi k_i)^{-1}$ are the FT temperatures of the two CFTs which are associated to the $\phi_i$ circles in \cite{CCLP}.
 The FT temperature of each point in the lattice of CFT duals is
 given by
 \be
 T_{(m, n)}=\frac{1}{2\pi(mk_1+nk_2)},
 \ee
 where $m$ and $n$ are similar to the ones used in Eq.(\ref{mn}).
 Using the Cardy formula, one can obtain the statistical entropy of dual
 CFT as follows which is in precise agreement with the
 Bekenstein-Hawking entropy (\ref{BH}),
 \be
 S_{\rm mic}=\frac{\pi^2}{3}c_{(m,n)} T_{(m,n)}=S_{\rm BH}.\label{cTS}
 \ee
 \section{Consistency of boundary conditions}\label{app}
 In this section we give explicitly the form of charges associated
 to the time translation generator $\rd_t$ and the general diffeomorphism generators given in Eq.(\ref{revised3}).
 As mentioned in Eq.(\ref{Q}) the charge associated to the generator $\zeta$
 is given by
 \be
 Q_\zeta=\frac{1}{8\pi}\int_{\rd\Sigma}k_{\zeta}[h,g],\label{QQ}
 \ee
 in which
 \bea
 k_\zeta[h,g]&=&\frac{1}{2}[\zeta_\nu\nabla_\mu h-\zeta_\nu\nabla_\sigma h_\mu^{~\sigma}+\zeta_\sigma\nabla_\nu h_\mu^{~\sigma}+
 \frac{h}{2}\nabla_\nu\zeta_\mu\nn\\&-&h_\nu^{~\sigma}\nabla_\sigma\zeta_\mu
 +\frac{1}{2}h_{\nu\sigma}(\nabla_\mu\zeta^\sigma+\nabla^\sigma\zeta_\mu)]*(dx^\mu\wedge
 dx^\nu).
 \eea
 Given $k_{\rd_t}$ in Eq.(\ref{rdt}), one verifies that the generator of time translation $Q_{\rd_t}=0$.
 This is a necessary condition for consistency of our calculations. We had started with extremal solutions (\ref{met}) and
 $Q_{\rd_t}=0$ ensures that by the boundary conditions (\ref{h}) extremality will not be violated.
 Furthermore, since the component of $k_{\zeta}$ given in Eq.(\ref{rdzeta}), that contribute in $Q_\zeta$  is
 independent of $r$,  the charges $Q_\zeta$ are finite under
 the boundary condition (\ref{h}).

 The explicit form of $k_{\rd_t}$  is given by,\footnote{Our convention is $\epsilon_{tr\phi_1\phi_2\theta}=1$.}
  \bea
 k_{\rd_t}&=&{r\over A^2B_1B_2}d\bigg[(10B_1B_2^{~2}Ck_1k_2-6AB_1B_2)(\lambda_1d\phi_2+\lambda_2
 d\phi_1)\nn\\
 &+&(A^2B_1+A^2B_2C-AB_1B_2k_2^{~2})(\frac{\rd^2\lambda_1}{\rd\phi_2^2}d\phi_2+\frac{\rd^2\lambda_2}{\rd\phi_2^2}d\phi_1)\nn\\
 &+&(A^2B_2-AB_1B_2k_1^{~2})(\frac{\rd^2\lambda_1}{\rd\phi_1^2}d\phi_2+\frac{\rd^2\lambda_2}{\rd\phi_1^2}d\phi_1)\nn\\
 &-&2(A^2B_2C+AB_1B_2k_1k_2)(\frac{\rd^2\lambda_1}{\rd\phi_1\rd\phi_2}d\phi_2+\frac{\rd^2\lambda_2}{\rd\phi_1\rd\phi_2}d\phi_1)\nn\\
 &+&4B_1B_2(k_2^{~2}B_2\lambda_1d\phi_2+k_1^{~2}B_1\lambda_2d\phi_1)\nn\\
 &+&6B_1B_2(k_1^{~2}B_1\lambda_1d\phi_2+k_2^{~2}B_2\lambda_2d\phi_1)\nn\\
 &+&2B_1B_2^{~2}C(k_1^{~2}\lambda_2d\phi_2+k_2^{~2}\lambda_1d\phi_1)
 +2B_1B_2^{~2}k_1k_2(\lambda_2d\phi_2+\lambda_1d\phi_1)\nn\\
 &+&2B_1B_2^{~2}C^2(3k_1^{~2}\lambda_1d\phi_2+2k_1^{~2}\lambda_2d\phi_1+k_1k_2\lambda_1d\phi_1)\bigg]_{\rd \Sigma}\wedge
 d\theta+k_{\rd_t}^\perp,
 \label{rdt}
 \eea
 where $k_{\rd_t}^\perp$ includes terms with components transverse to the boundary. So,
 \be
 \int_{\rd\Sigma}k_{\rd_t}^\perp=0.
 \ee
 Likewise,
 \bea
 k_\zeta&=&{1\over 2A^2}\bigg[2(\frac{\rd\lambda_1}{\rd{\phi_1}}\lambda_2-\lambda_1\frac{\rd\lambda_2}{\rd\phi_1}+\lambda_1\frac{\rd\lambda_2}{\rd\phi_2}C)B_2(k_1C+k_2)\nn\\
 &-&2\frac{\rd\lambda_1}{\rd\phi_2}\lambda_2(k_1B_1+k_2B_2C)+
 k_2A\frac{\rd^2\lambda_1}{\rd\phi_1^2}(\frac{\rd\lambda_2}{\rd\phi_2}+\frac{\rd\lambda_1}{\rd\phi_2})\nn\\
 &+&2A\lambda_1(k_1\frac{\rd^3\lambda_2}{\rd\phi_1^2\rd\phi_2}+k_2\frac{\rd^3\lambda_1}{\rd\phi_1^2\rd\phi_2})
 +k_1A\frac{\rd\lambda_1}{\rd\phi_1}(\frac{\rd^2\lambda_1}{\rd\phi_1^2}+\frac{\rd^2\lambda_2}{\rd\phi_1\rd\phi_2})\nn\\
 &+&k_2A(\frac{\rd^2\lambda_1}{\rd\phi_2^2}\frac{\rd\lambda_2}{\rd\phi_2}+\frac{\rd\lambda_1}{\rd\phi_2}\frac{\rd^2\lambda_2}{\rd\phi_1\rd\phi_2})\nn\\
 &+&k_1A\frac{\rd\lambda_2}{\rd\phi_1}(\frac{\rd^2\lambda_1}{\rd\phi_1\rd\phi_2}+\frac{\rd^2\lambda_2}{\rd\phi_2^2})
 +2k_1(B_1\lambda_1\frac{\rd\lambda_2}{\rd\phi_2}-B_2C^2\lambda_2\frac{\rd\lambda_1}{\rd\phi_2})\nn\\
 &+&2k_1A(\lambda_1\frac{\rd^3\lambda_1}{\rd\phi_1^3}+\lambda_2\frac{\rd^3\lambda_1}{\rd\phi_1^2\rd\phi_2}+
 \lambda_2\frac{\rd^3\lambda_2}{\rd\phi_1\rd\phi_2^2})\nn\\
 &+&2k_2A(\lambda_2\frac{\rd^3\lambda_2}{\rd\phi_2^3}+
 \lambda_1\frac{\rd^3\lambda_2}{\rd\phi_1\rd\phi_2^2}+\lambda_2\frac{\rd^3\lambda_1}{\rd\phi_1\rd\phi^2_2})\bigg]
 d\phi_1\wedge d\phi_2\wedge d\theta\nn\\
 &+&k_\zeta^\perp.
 \label{rdzeta}\eea

  %----------------------------------------------------------------------------------------------------------------
 \section*{Summary }\label{d}
 We showed  that the moduli space of the near horizon metric of five
 dimensional extremal black holes that are asymptotically flat or
 AdS has an SL(2,${\mathbb Z}$) modular symmetry which leaves the
 Bekenstein-Hawking entropy invariant. By using the Brown-Henneaux approach,
 we obtained  CFTs associated to the points in moduli space which are identified
 by the modular group.
 In this way we realized a lattice of chiral CFTs dual to any such
 geometry. The  microscopic entropy of each lattice point is
 equivalent to the Bekenstein-Hawking entropy.

 \section*{Acknowledgement} We would like to thank R. Fareghbal for valuable discussions.
 %---------------------------------------------------------------------------------------------------------------
 %---------------------------------------------------------------------------------------------------------------

 %---------------------------------------------------------------------------------------------------------------

%-------------------------------------------------------------------------------------------------------------------------
 %\newpage


\begin{thebibliography}{99}
 \bibitem{SV}A. Strominger and C. Vafa, {\it Microscopic Origin of the Bekenstein-Hawking Entropy},
 Phys. Lett. B379 (1996) 99–104, [hep-th/9601029].
 \bibitem{GHSS}M. Guica, T. Hartman, W. Song and A. Strominger, {\it The Kerr/CFT correspondence}, [arXiv:0809.4266].
 \bibitem{BH}J. D. Brown and M. Henneaux, {\it Central Charges in the Canonical Realization of
 Asymptotic Symmetries: An Example from Three-Dimensional Gravity},
 Commun. Math. Phys. 104 (1986) 207–226.
 \bibitem{strominger}A. Strominger, {\it  Black Hole Entropy from Near-Horizon Microstates
 }, JHEP 02 (1998) 009, [hep-th/9712251].
 \bibitem{C}S. Carlip, {\it Black hole entropy from conformal field theory in any dimension}, Phys. Rev.
 Lett. 82 (1999) 2828, [hep-th/9812013].
 \bibitem{S}S.N. Solodukhin, {\it Conformal description of horizon's states}, Phys. Lett. B454 (1999)
 213, [hep-th/9812056].
 \bibitem{P}M.I. Park, {\it Hamiltonian dynamics of bounded spacetime and black hole entropy: Canonical
 method}, Nucl. Phys. B634 (2002) 339, [hep-th/0111224].
 \bibitem{LMP}H. L\"{u}, J. Mei and C.N. Pope, {\it Kerr-AdS/CFT Correspondence in Diverse Dimensions},
 [arXiv:0811.2225].
 \bibitem{ITW} H. Isono, T.S. Tai and W.Y. Wen,
 {\it  Kerr/CFT correspondence and five-dimensional BMPV black
 holes}, [arXiv:0812.4440].
 \bibitem{AOT}T. Azeyanagi, N. Ogawa and S. Terashima, {\it Holographic Duals of Kaluza-Klein Black
 Holes}, [arXiv:0811.4177].
 \bibitem{CCLP}David D.K. Chow, M. Cvetic,  H. L\"{u}  and C.N. Pope, {\it Extremal Black Hole/CFT Correspondence in (Gauged)
 Supergravities}, [arXiv:0812.2918].
 \bibitem{N}Y. Nakayama, {\it Emerging AdS from Extremaly Rotating NS5-branes},
 [arXiv:0812.2234].
 \bibitem{AOT1}T. Azeyanagi, N. Ogawa and S. Terashima, {\it The Kerr/CFT Correspondence and String Theory}, [arXix:0812.4883].
 \bibitem{KLR}H. Kunduri, J. Lucietti and H.S. Reall, {\it Near-horizon symmetries of extremal black
 holes}, Class. Quant. Grav.24 4269-4190, 2007 [arXiv:0705.4214].
 \bibitem{BB}G. Barnich and F. Brandt, {\it Covariant theory of asymptotic symmetries, conservation
 laws and central charges}, Nucl. Phys. B633 (2002) 3–82,
 [hep-th/0111246].
 \bibitem{BC}G. Barnich and G. Compere, {\it Surface charge algebra in gauge theories and
 thermodynamic integrability}, J. Math. Phys. 49 (2008) 042901, [arXiv:0708.2378].
 \bibitem{FT}V.P. Frolov and K.S. Thorne, {\it Renormalized stress-energy tensor near the horizon of a
 slowly evolving, rotating black hole}, Phys. Rev. D39 (1989) 2125.
 \bibitem{HMNS}T. Hartman, K. Murata, T. Nishioka, and A.
 Strominger, {\it CFT Duals for extreme Black Holes},
 [arXiv:0811.4393].
 \bibitem{LSNHBR} F. Loran, H. Soltanpanahi, {\it Near the horizon of 5D black rings}, [arXiv:0810.2620]

 \end{thebibliography}
\end{document}